\documentclass[12 pt]{article}
\usepackage{amsmath, amsfonts, amssymb}
\usepackage{graphicx}
\usepackage{indentfirst}
\usepackage{threeparttable}
\usepackage{url}
\begin{document}

\title{\LARGE \bf Spectral Analysis of the Orbital Dynamics of Globular Clusters in the Central Region of the Milky Way}
\author{\bf A.~T.~Bajkova\thanks{E-mail: anisabajkova@mail.ru},
A.~A.~Smirnov 
and
V.~V.~Bobylev
}
\date{\it  \small  $^1$ Pulkovo Astronomical Observatory, \\
St.-Petersburg 196140, Russia}

\maketitle




\begin{abstract}
A new method for determining the nature of the orbital motion (chaotic or regular) of globular clusters in the central region of the Galaxy with a radius of 3.5 kpc, which are most affected by the bar, is proposed. The method is based on calculating the orbital power spectrum as a function of time and calculating the entropy of the power spectrum as a measure of orbital chaos. The sample includes 45 globular clusters. To form the 6D phase space required for integrating the orbits, the most accurate astrometric data to date from the Gaia satellite (Vasiliev \& Baumgardt, 2021) were used, as well as new refined average distances (Baumgardt \& Vasiliev, 2021). Orbits of globular clusters are obtained in a non-axisymmetric potential with a bar in the form of a triaxial ellipsoid embedded in an axisymmetric potential, traditionally used by us to construct orbits of globular clusters, described in detail in the paper by Bajkova et al. (2023a). The following, most realistic, bar parameters are adopted: mass $10^{10} M_\odot$, length of the major semi-axis 5 kpc, angle of rotation of the bar axis 25$^o$, angular velocity of rotation 40 km s$^{-1}$ kpc$^{-1}$. A list of 23 globular clusters with regular dynamics and 22 globular clusters with chaotic dynamics is determined. The correlation of the obtained classification of globular clusters with the classification obtained by us using other methods in the work of Bajkova et al. (2024a) was determined.
\end{abstract}


\section{Introduction}
This work is a continuation of a series of works by the authors (Bajkova \& Bobylev, 2022; Bajkova et al., 2023a; Bajkova et al., 2023b; Smirnov et al., 2023; Bajkova et al., 2023c; Smirnov et al., 2024; Bajkova et al., 2024a; Bajkova et al., 2024b) devoted to the study of the orbital dynamics of globular clusters (GCs). Thus, in work by Bajkova \& Bobylev (2022) a catalog of orbits of 152 galactic globular clusters  based on the latest astrometric data from the Gaia satellite (Gaia EDR3) (Vasiliev \& Baumgardt, 2021), as well as new refined average distances (Baumgardt \& Vasiliev, 2021) is presented. In work by Bajkova et al. (2023a) an analysis (based on the same data) of the influence of the galactic bar on the orbital motion of globular clusters in the central region of the Galaxy  was performed. For this task, 45 globular clusters in the central galactic region with a radius of 3.5 kpc were selected (a list of these GCs is given below in the table with the results of this work). The orbits of the globular clusters were obtained both in an axisymmetric potential and in a potential including a bar model in the form of a triaxial ellipsoid. In this case, the mass, rotation velocity and dimensions of the bar were varied. A comparison of such orbital parameters as apocentric and pericentric distances, eccentricity and maximum distance from the galactic plane was made.

The second stage of the research aimed at studying the influence of the bar on the orbital motion of globular clusters was devoted to the problem of identifying objects captured by the bar using spectral dynamics methods (Bajkova et al., 2023b; Smirnov et al., 2023; Bajkova et al., 2023c; Smirnov et al., 2024).

The third stage of the research was devoted to the analysis of the regular/chaotic nature of the orbits of all 45 selected GCs using various methods (Bajkova et al., 2024a; Bajkova et al., 2024b). Namely, 1) methods for calculating the maximum characteristic Lyapunov indices (in the classical version and in the version with renormalization of the "shadow" orbit corresponding to the perturbed initial phase points relative to the "reference" orbit with given initial phase points), 2) Poincare sections, 3) a frequency method based on calculating fundamental frequencies, as well as 4) a visual assessment based on images of the reference and shadow orbits. In this case, the model of the bar was adopted as a model of an elongated triaxial ellipsoid with the most probable parameters known from the literature (see, for example, Palous et al. (1993), Sanders et al.(2019)): mass $10^{10} M_\odot$, length of the major semiaxis of 5 kpc, angle of inclination to the galactic axis $X$ 25$^o$, rotation speed of 40 km s$^{-1}$ kpc$^{-1}$.

Since the GCs in the central region of the Galaxy are subject to the greatest influence from the elongated rotating bar, the question of the nature of the orbital motion of the GCs - regular or chaotic - is of great interest. For example, in Machado \& Manos (2016) it is shown that
the main share of chaotic orbits should be precisely in the bar region.

This work is essentially a continuation of the third stage, devoted to the study of the chaotic dynamics of GCs in the central region of the Galaxy. In Bajkova et al. (2024a) we investigated the orbital dynamics of GCs only in a potential with a bar, in Bajkova et al. (2014b) a comparison of the orbital dynamics of GCs in an axisymmetric potential and in a non-axisymmetric potential is given in order to determine the effect of the bar on the degree of chaos of GC orbits. If in previous works we used well-known methods for analyzing the regularity/chaos of orbits, then in this work we propose a new approach based on calculating the power spectrum of the reference and shadow orbits of GCs and using the entropy measure to analyze the spectra as a measure of orbit chaos. For additional control of the obtained results, we also use the Poincare section method, the frequency method, and the method of visually assessing the degree of divergence of the reference and shadow orbits over time.

The work is structured as follows. The second section gives a brief description of the adopted potential models -- an axisymmetric potential and a non-axisymmetric potential including a bar. The third section provides links to the astrometric data used, as well as to the method for forming the GC sample. The fourth section describes the proposed method for assessing the regularity/chaoticity of motion, based on the spectral analysis of orbits and the calculation of the spectrum entropy as a measure of the chaoticity of orbits, as well as known methods -- the Poincare section method and the frequency method. The fifth section analyzes the results obtained. The Conclusion formulates the main findings of the work.

\section{Galactic Potential Model}

\subsection{Axisymmetric potential}

The axisymmetric gravitational potential of the Galaxy, traditionally used by us (see, for example, Bajkova \& Bobylev (2022)) for integrating the GC orbits, is represented as the sum of three components~--- the central spherical bulge $\Phi_b(r)$, the disk $\Phi_d(R,Z)$ and the massive spherical halo of dark matter $\Phi_h(r)$:
 \begin{equation}
 \begin{array}{lll}
  \Phi(R,Z)=\Phi_b(r)+\Phi_d(R,Z)+\Phi_h(r).
 \label{pot}
 \end{array}
 \end{equation}
Here we use a cylindrical coordinate system ($R,\psi,Z$) with
the origin at the center of the Galaxy. In a rectangular coordinate system $(X,Y,Z)$ with the origin at the center of the Galaxy, the distance to a star (spherical radius) will be equal to
$r^2=X^2+Y^2+Z^2=R^2+Z^2$, with the $X$ axis directed from the Sun to the galactic center, the $Y$ axis perpendicular to the $X$ axis in the direction of the Galaxy's rotation, and the $Z$ axis perpendicular to the galactic plane $(X,Y)$ in the direction of the north galactic pole. The gravitational potential is expressed in units of 100 km$^2$ s$^{-2}$, distances~--- in kpc, masses~--- in units of the galactic mass $M_{gal}=2.325\times 10^7 M_\odot$,
corresponding to the gravitational constant $G=1$.

The axisymmetric potentials of the bulge $\Phi_b(r(R,Z))$ and the disk $\Phi_d(r(R,Z))$ are represented in the form proposed by Miyamoto \& Nagai (1975):
 \begin{equation}
  \Phi_b(r)=-\frac{M_b}{(r^2+b_b^2)^{1/2}},
  \label{bulge}
 \end{equation}
 \begin{equation}
 \Phi_d(R,Z)=-\frac{M_d}{\Biggl[R^2+\Bigl(a_d+\sqrt{Z^2+b_d^2}\Bigr)^2\Biggr]^{1/2}},
 \label{disk}
\end{equation}
where $M_b, M_d$~ are component masses, $b_b, a_d, b_d$~ are component scale parameters in kpc. The halo component (NFW) is represented according to the work~ Navarro et al. (1997):
 \begin{equation}
  \Phi_h(r)=-\frac{M_h}{r} \ln {\Biggl(1+\frac{r}{a_h}\Biggr)}.
 \label{halo-III}
 \end{equation}

Table~1 presents the values of the parametrers of the Galactic potential model (2)--(4), which were found by
Bajkova \& Bobylev (2016) using the Galactic rotation curve of Bhattacharjee et al (2014), constructed based on objects located
at distances $R$ up to $\sim200$~kpc. Note that when constructing this Galactic rotation curve, the
following values of the local parameters were used: $R_\odot=8.3$~kpc and $V_\odot=244$~km s$^{-1}$. In
Bajkova \& Bobylev (2016) the model (2)--(4) is designated as model~III. The adopted potential model is the best among the six models considered in Bajkova \& Bobylev (2017), since it provides the smallest discrepancy between the data and the model rotation curve.

\subsection{Bar model}

The model of a three-axis ellipsoid was chosen as the central bar potential according to Palous et al. (1993):
\begin{equation}
  \Phi_{bar} = -\frac{M_{bar}}{(q_b^2+X^2+[Ya/b]^2+[Za/c]^2)^{1/2}},
\label{bar}
\end{equation}
where $X=R\cos\vartheta, Y=R\sin\vartheta$, $a, b, c$~ are three
semi-axes of the bar, $q_b$~ is scale parameter of the bar (length of the largest semi-axis of the bar);
$\vartheta=\theta-\Omega_{b}t-\theta_{b}$, $tg(\theta)=Y/X$,
$\Omega_{b}$~ is circular velocity of the bar, $t$~ is integration time, $\theta_{b}$~is orientation angle of the bar relative to the galactic axes $X,Y$, measured from the line
connecting the Sun and the center of the Galaxy (axis $X$) to the major axis of the bar in the direction of rotation of the Galaxy.

Based on information in numerous literature, in particular, in Palous et al. (1993), the following were used as bar parameters: $M_{bar}=430\times M_{gal}$, $\Omega_{b}=40$~km s$^{-1}$ kpc$^{-1}$, $q_b=5$ kpc, $\theta_{b}=25^o$. The adopted bar parameters are listed in table~1.

 {\begin{table}[t]                                    
 \caption[]
 {\small\baselineskip=1.0ex Values of the parameters of the galactic potential model, $M_{gal}=2.325\times 10^7 M_\odot$}
 \begin{center}\begin{tabular}{|c|r|}\hline
 $M_b$ &   443 M$_{gal}$ \\
 $M_d$ &  2798 M$_{gal}$ \\
 $M_h$ & 12474 M$_{gal}$ \\
 $b_b$ & 0.2672 kpc  \\
 $a_d$ &   4.40 kpc  \\
 $b_d$ & 0.3084 kpc  \\
 $a_h$ &    7.7 kpc  \\
\hline\hline
 $\Omega_b$ & 40 km s$^{-1}$ kpc$^{-1}$ \\
 $q_b$     &  5.0 kpc  \\
 $\theta_{b}$ &  $25^o$   \\\hline
 $a/b$ & 2.38  \\
 $a/c$ & 3.03  \\
    \hline
 \end{tabular}\end{center}\end{table}}

To integrate the equations of motion, we used the fourth-order Runge-Kutta algorithm.

The value of the peculiar velocity of the Sun relative to the local
standard of rest was taken to be equal to
$(u_\odot,v_\odot,w_\odot)=(11.1,12.2,7.3)\pm(0.7,0.5,0.4)$~km s$^{-1}$
according to the work of Sch\"onrich et al. (2010). The elevation of the Sun above the plane of the Galaxy was taken to be equal to 16 pc in accordance with the work of Bobylev \& Bajkova (2016).

For comparison, Fig.~1 shows the obtained model rotation curves: an axisymmetric potential (black line) and a potential with a bar (red line).

\begin{figure*}
\hskip 3cm
   \includegraphics[width=0.5\textwidth,angle=-90]{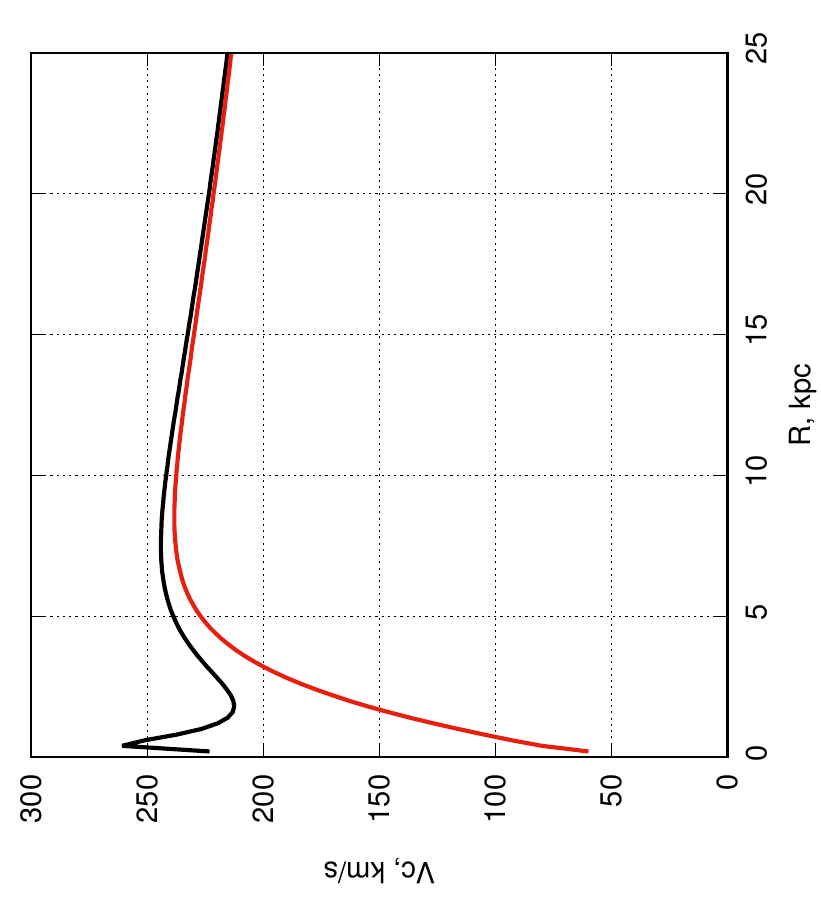}
\caption{\small Rotation curve of the Galaxy with an axisymmetric potential without a bar (black line) and a non-axisymmetric potential including a bar (red line).}
 \label{fDD}
\end{figure*}

\section{Data}

The data on the proper motions of GCs are taken from the new catalogue by Vasiliev \& Baumgardt (2021), compiled on the basis of Gaia EDR3 observations. The GC coordinates and radial velocities are taken from Vasiliev (2019). The average distances to globular clusters are taken from Baumgardt \& Vasiliev (2021). A comparative analysis of the new data on proper motions and distances with previous versions of the catalogues is given, for example, in Bajkova \& Bobylev (2022).

The GCs catalog at our disposal Bajkova \& Bobylev (2022) contains 152 objects. The selection of globular clusters from this set, belonging to the bulge/bar region, was made in accordance with a purely geometric criterion, considered in Massari et al. (2019), and also used by us in Bajkova et al. (2020). It is very simple and consists of selecting GCs, the apocentric distance of whose orbits does not exceed the bulge radius, which is usually taken to be 3.5 kpc. In this case, the orbits are calculated in an axisymmetric potential. The full list of 45 objects in our sample is listed in Table~2, which presents the results of the analysis of the chaoticity/regularity of the GC orbit (the first column gives the serial number of the GC, the second -- the name of the GC).
\label{lab:text1}

\section{Methods of analysis of regularity / chaoticity of orbital dynamics of GCs}

\subsection{Poincare sections}

One of the methods for determining the nature of the motion (regular or chaotic) is the analysis of Poincare sections Morbidelli (2014). The algorithm used by us to construct the mappings is as follows:

1. We consider the phase space $(X,Y,V_x,V_y)$.

2. We eliminate $V_y$ using the conservation law of the generalized energy integral (Jacobi integral) and move to the space $(X,Y,V_x)$.

3. We define the plane $Y=0$, the intersection points with the orbit are designated on the plane $(X,V_x)$. We take only those points where $V_y>0$.

The phase space $(Y,Z,V_y,V_z)$ or $(R,Z,V_R,V_z)$ can be considered similarly. Then the Poincare sections will be reflected on the plane $(Y,V_y)$ or $(R,V_R)$, respectively.

If the intersection points of the plane add up to a continuous smooth line (or several separated lines), then the motion is considered regular. In the case of chaotic motion, instead of being located on a smooth curve, the points fill a two-dimensional region of phase space, and sometimes there is an effect of sticking of points to the boundaries of islands corresponding to ordered motion (Morbidelli, 2014).

In this paper we present the Poincare sections obtained in Bajkova et al. (2024a).

\subsection{Frequency method}

The described method of studying the regularity/chaoticity of orbits is associated with the use of orbital frequencies (Nieuwmunster et al., 2024; Valluri et al., 2010)(see Section 3.1 in the last paper). The authors of these works showed that it is possible to measure the stochasticity of an orbit based on the shift of fundamental frequencies determined over two consecutive time intervals. For each frequency component $f_i$, a parameter called the frequency drift is calculated:
\begin{equation}
\label{freq}
\lg(\Delta f_i)=\lg|\frac{\Omega_i(t_1)-\Omega_i(t_2)}{\Omega_i(t_1)}|,
\end{equation}
where $i$ defines the frequency component in Cartesian coordinates
(i.e. $\lg(\Delta f_x), \lg(\Delta f_y)$ and $\lg(\Delta f_z)$). Then the largest value of these three frequency drift parameters $\lg(\Delta f_i)$ is assigned to the frequency drift parameter $\lg(\Delta f)$. The higher the value of $\lg(\Delta f)$, the more chaotic the orbit. However, as shown in Valluri et al. (2010), the accuracy of the frequency analysis requires at least 20 oscillation periods to avoid classification errors. In order to achieve high accuracy, we took an integration time of 120 billion years, almost an order of magnitude greater than the age of the Universe.

In this work, we used the classification results obtained by the frequency method, given in Bajkova et al. (2024a).

\subsection{A new method on the basis of spectral analysis}

In our case, the spectral analysis of orbits is based on the calculation of the modulus of the discrete Fourier transform (DFT) of a uniform time series of radial distances of orbit points from the center of the Galaxy $r_n$, calculated from their $X, Y, Z$ galactic coordinates $X(t_n), Y(t_n), Z(t_n)$ as functions of time: $r(t_n)=\sqrt{X(t_n)^2+Y(t_n)^2+Z(t_n)^2}$, where $n=0,...,N-1$ ($N$ is the length of the series).

Thus, the formula for the DFT  module (power spectrum) of the $r_n$ sequence will look like this:
\begin{equation}
\label{Equ1}
R_k=|\frac{1}{N} \sum_{n=0}^{N-1} r_n\exp{(-\jmath\frac{2\pi\times n\times k}{N})}|,~~~ k=0,..., N-1.
\end{equation}
In this case, the length of the series is chosen to be $N=2^\alpha$, where $\alpha$ is an integer, >0, so that the fast Fourier transform (FFT) algorithm can be used to calculate the DFT. The required length of the series is achieved by supplementing the real series with zeros.

In our case, the length of the real sequences is 120,000, since we integrate the orbits back 120 billion years with an integration interval of 1 million years. Before calculating the DFT, we pre-center the coordinate series (i.e. get rid of the constant component), then supplement the resulting sequence $r_n$ with zero counts at $n>120,000$ until the length of the entire analyzed sequence reaches $N=262144=2^{18}$. Note that supplementing the original sequence with zeros is also useful in terms of increasing the accuracy of the coordinates of the spectral components. Since the interval between the sequence counts in time is $\Delta_t=0.001$ billion years, the analyzed frequency range, which is a periodic function, is $F=1/\Delta_t=1000$ Gyr$^{-1}$. The frequency discrete is $\Delta_F= F/N \approx 0.03815$ Gyr$^{-1}$. In the following, for convenience, we will indicate on the graphs not physical frequencies, but the numbers of samples $k$ (or $K$) of the discrete Fourier transform (\ref{Equ1}). The transition from $k$ to the physical frequency can be made using the formula
$f=k\times\Delta_F\approx k\times 0.003815$. Then the obtained power spectrum of the GC orbit is normalized so that the maximum value is equal to unit.

The decision on the nature of the orbital dynamics of the GC is determined by calculating the Shannon entropy of the normalized power spectrum $R_k$ as a measure of chaos (Chumak, 2011):
\begin{equation}
\label{Equ2}
E_R=-\frac{1}{M}\sum_{k=0}^{N-1} R_k \ln(R_k),
\end{equation}
where $M$ is a scale factor that is introduced for the convenience of presenting numerical results.

Obviously, the higher the entropy value, the higher the degree of orbital chaos. Since we analyze both reference orbits and shadow orbits obtained by perturbing the initial phase point (see the next section), we should also pay attention not only to the entropy value, but also to the difference in the entropy values of the reference and perturbed orbit spectra. Obviously, in the case of a regular orbit, this difference should be small enough by analogy with the Lyapunov indices. In addition, we note that by analogy with the Lyapunov indices, we take the orbital integration time to be large enough. In our case (see the next section), this time is, as in the case of the frequency method, 120 billion years, which, as already mentioned, is almost an order of magnitude greater than the age of the Universe.

\section{Results of orbital analysis, their classification and comparison of methods}

The obtained results of classification of the orbital dynamics of 45 GC in the center of the Galaxy using a new method based on spectral analysis (7) and the use of the entropy measure (8) as a measure of chaos, as well as the known methods of Poincare sections, the frequency method and visual analysis of orbits in projection onto the galactic $(X-Y)$ plane are reflected in Table 2 and in Figs.~2 and ~3 (see figure captions).

The proposed method was applied both for reference orbits, designated by us (see Fig. 3) as (E), according to the initial data, and for shadow orbits, designated as (F), with a perturbation of the initial phase point, as in Bajkova et al. (2024a), as follows: $X_1=x_0+X_0\times 0.00001,~Y_1=Y_0+Y_0\times 0.00001,~Z_1=Z_0+Z_0\times 0.00001$.

The integration of the orbits was carried out, as already noted above, for 120 billion years backward.
Fig.~2 shows from left to right: $X-Y$ projections of the orbits constructed on the time interval $[-11,-12]$ billion years and taken from Bajkova et al. (2024a); radial values of the initial (reference) and perturbed (shadow) orbits depending on time ($r(t)$) (the reference orbits are shown in yellow, the shadow orbits in purple); normalized power spectra of the radial values of the reference and shadow orbits as functions of time on the time interval $[0, -120]$ billion years, calculated in this work and shown in black and red, respectively; Poincare sections $X-V_x$ from Bajkova et al. (2024a); normalized power spectra of the Poincare sections as functions of the coordinates $X$ and $V_x$ on time, also calculated in this work and shown in purple and yellow, respectively; illustration of the frequency method from work Bajkova et al. (2024a) (the power spectrum of the first half of the time sequence is shown in red, the second half in black).

The most obvious illustration of the discrepancy between the reference and shadow phase points are the left columns 1 and 2 of Fig.~2, which show the reference and shadow orbits for each GC in the order (from top to bottom) as they are given in Table 2. The first column shows the $X-Y$ projections of the orbits constructed in the rotating bar system on the time interval $[-11,-12]$ billion years. The second column shows the radial values of the orbit $r(t)$ on the interval $[0,-12]$ billion years, comparable with the age of both the GCs and the Universe.
In these graphs, the reference orbits are shown in yellow, and the shadow orbits in purple. It can be seen that for many objects, only purple is present on the graphs. This means that the shadow orbit practically coincides with the reference one (the yellow lines are covered by purple ones). Such objects include GCs with regular orbits. On the graphs of GCs with chaotic orbits, both purple and yellow lines are visible, which allows us to qualitatively judge the degree of chaos of the orbits.

Let us note right away that the normalized power spectra of the radial values of the reference and shadow orbits, given in the third column of Fig.~2, have the character of line spectra for GCs with regular dynamics and wide spectra for GCs with chaotic dynamics, which leads to a noticeable increase in entropy in the latter case. We also observe the broadening of the spectra for GCs with chaotic dynamics in columns 5 and 6 of Fig.~2.

Let us turn to Table~2. The first column contains the ordinal numbers of the globular clusters in our sample, and the second column contains the names of the globular clusters. The calculated values of the entropy of the power spectra of the reference orbits, shadow orbits, and the differences in the entropy of the power spectra of the reference and shadow orbits, pre-scaled with a coefficient of $M=10000$ (see expression (8)) in order to obtain relatively small numbers, are given in the third, fourth, and fifth columns, respectively. (Note that scaling the entropy values for all orbits with the same coefficient does not affect the decision on the chaotic/regular nature of the orbits, since the comparison problem is being solved.)

The classification of orbits based on the chaotic (C) or regular (R) characteristics as a result of applying the new method, designated in the table as (1), is given in the sixth column. The classifications of orbits based on the frequency method (2), visual analysis (3), and the Poincare section method (4), taken from Bajkova et al. (2024a), are given for comparison with the new classification in the seventh, eighth, and ninth columns, respectively.

Let us note right away that the classification results obtained using the four methods listed are very close to each other. Thus, the correlation coefficient of the new method (1) with methods (2), (3), (4) is 0.825. At the same time, the GCs Terzan3, NGC 6304, NGC 6316 changed their classification to the opposite one, given in the work Bajkova et al. (2024a). We believe that the GCs Terzan3 and NGC 6316 can be classified as weakly chaotic. But this issue of classification requires a separate study, which is planned for the future.

The results of applying the new method are also clearly demonstrated in Fig.~3. Thus, the figure ({\bf a}) shows the scaled values of the entropy of the normalized power spectra for 45 GCs (ordinal number on the abscissa axis) for the reference orbits (red dots) and shadow orbits (black triangles); the figure ({\bf b}) shows the histograms for the entropy of the normalized power spectra of the reference (green) and shadow (purple) orbits of 45 GCs; the figure ({\bf c}) shows a diagram comparing the values of the entropy of the normalized power spectra of the reference (E) (abscissa axis) and shadow (F) (ordinate axis) orbits.

As can be seen from the figure ({\bf a}) and also from Table~2 (columns 3-5), globular clusters with entropy values $<0.02$ have practically equal values for the reference and shadow orbits, i.e. perturbation of the initial point practically did not lead to a change in entropy. This also follows from the figure ({\bf c}). With increasing entropy, the difference between the entropy values for the reference and shadow orbits grows. Therefore, globular clusters with entropy values $<0.02$, which was taken as the threshold under the adopted conditions of orbit integration and the value of the scale factor $M$, we classify as globular clusters with regular dynamics (R), and the rest as globular clusters with chaotic dynamics (C). As follows from the histogram of the entropy distribution for the reference and shadow orbits (Figure ({\bf b})), as a result of the perturbation of the initial point, the entropy value for some globular clusters decreased and became <0.02, but when making a decision on the nature of the dynamics, we also look at the entropy increment. For GCs with regular dynamics, it should be sufficiently small (as a rule, it is an order of magnitude or more smaller than for GCs with chaotic dynamics; see column 5 of Table~2).
Essentially, what has been said represents an algorithm for making a decision on the dynamic nature of a GC (R) or (C) as a result of applying the proposed method based on the spectral analysis of the radial values of the orbits as a function of time over an interval of 120 billion years and the calculation of the entropy of the obtained normalized spectra using the scale factor $M=10000$. As additional modeling has shown, with reasonable changes in the integration time and the scale factor $M$, only the threshold value of entropy changes, which has virtually no effect on the classification results. We consider the parameters adopted in this work to be optimal.

As a result of applying our method, we assigned the following globular clusters to the first group of objects with regular dynamics (R): NGC6266, Terzan4, Liller1, NGC6380, Terzan1, Terzan5, Terzan6, Terzan9, NGC6522, NGC6528, NGC6624, NGC6637, NGC6717, NGC6723, NGC6304, Pismi26, NGC6569, E456-78, NGC6540, Djorg2, NGC6171, NGC6539, NGC6553. The second group of objects with chaotic dynamics (C) includes globular clusters NGC6144, E452-11, NGC6273, NGC6293, NGC6342, NGC6355, Terzan2, BH229, NGC6401, Pal6, NGC6440, NGC6453, NGC6558, NGC6626, NGC6638, NGC6642, Terzan3, NGC6256, NGC6325, NGC6316, NGC6388, NGC6652. The first group contains 23 GCs, the second -- 22.

\begin{figure*}
\hskip -4cm
   \includegraphics[width=1.2\textwidth,angle=-90]{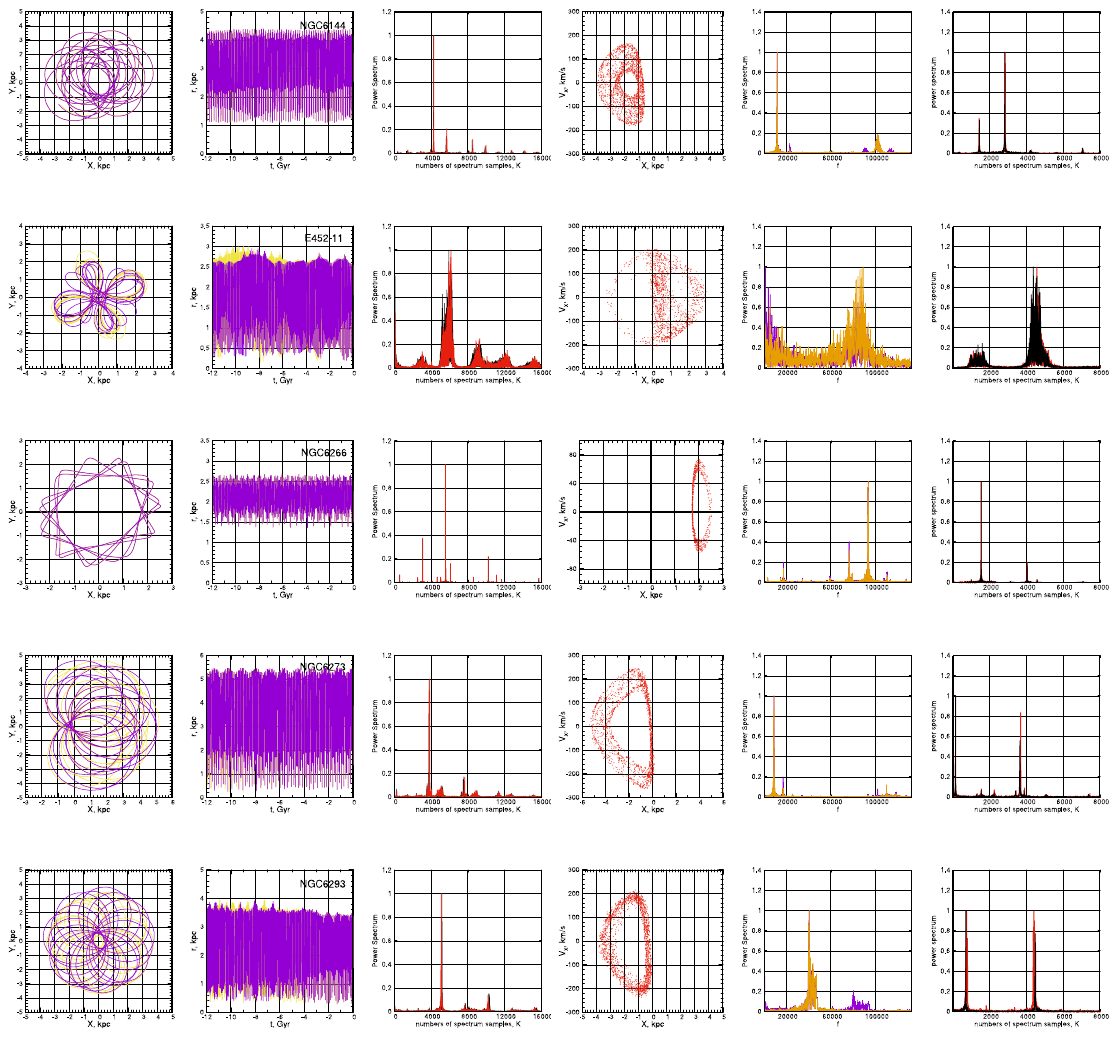}
  \vskip -1cm
\caption{\small From left to right: $X-Y$ projections of the orbits; radial values of reference (yellow) and shadow (purple) orbits as functions of time; normalized power spectra of the radial values jf thr reference  (black) and shadow (red) orbits as functions of time; $X-V_x$ Poincare sections; normalized power spectra of the Poincare sections coodinates $X$  (purple) and $V_x$ (yellow) coordinates as functions of time; illustration of the frequency method: the power spectrum of the first half of the time sequence (red) and of the second half (black). }
\end{figure*}

\begin{figure*}
\hskip -4cm
    \includegraphics[width=1.2\textwidth,angle=-90]{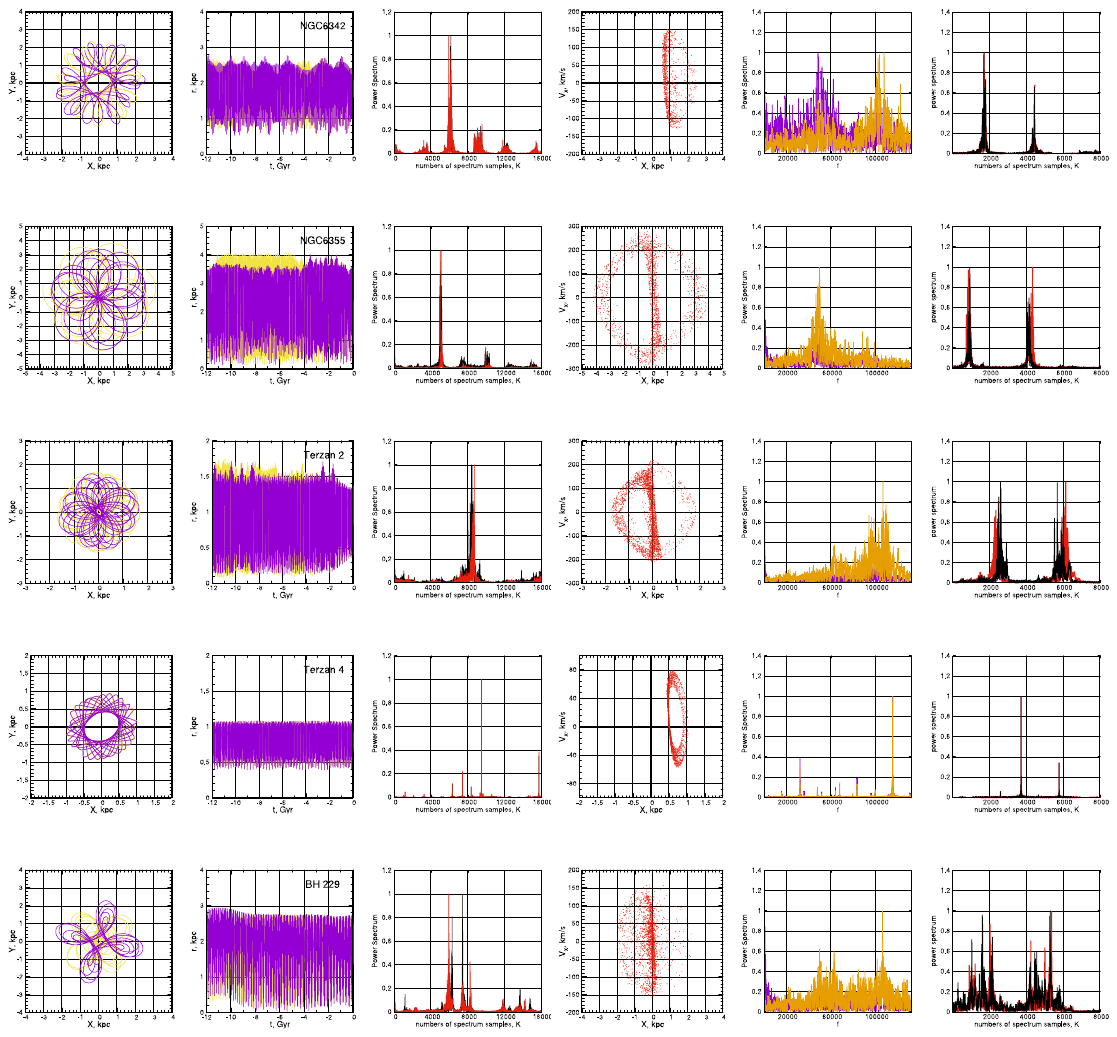}
\centerline{Figure 2:~{\small Continued.}}
 \label{fD}
\end{figure*}

\begin{figure*}
\hskip -4cm
    \includegraphics[width=1.2\textwidth,angle=-90]{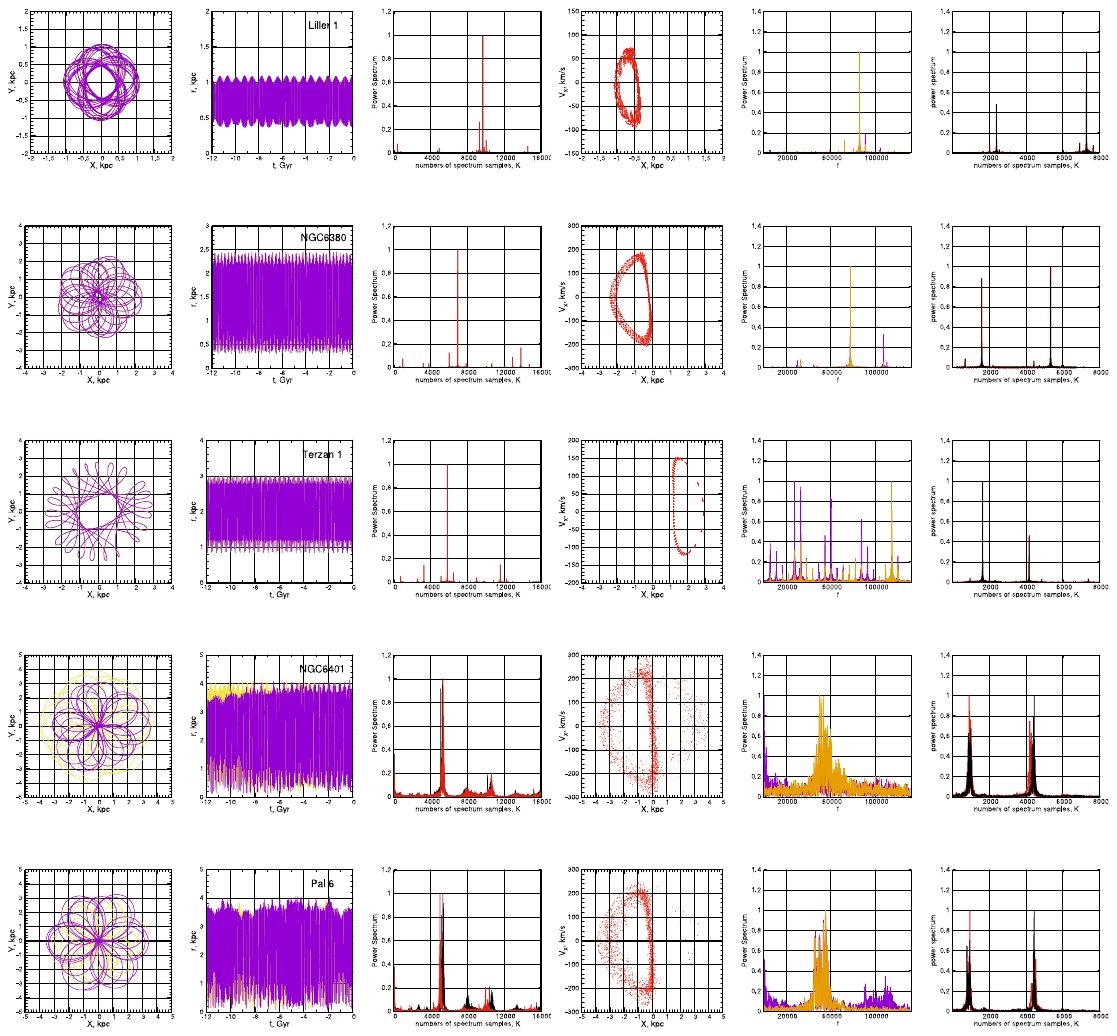}
\centerline{Figure 2:~{\small Continued.}}
 \label{fD}
\end{figure*}

\begin{figure*}
\hskip -4cm
    \includegraphics[width=1.2\textwidth,angle=-90]{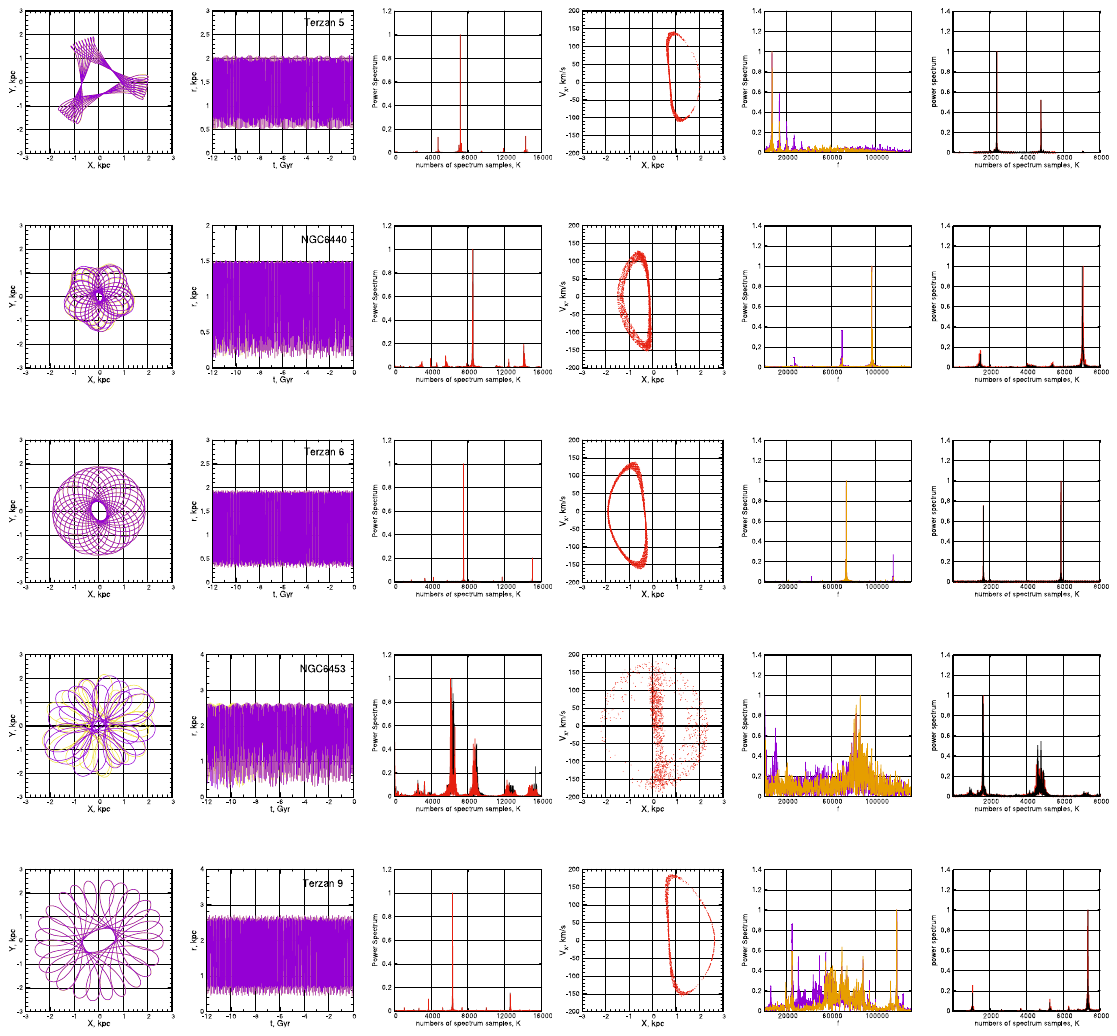}
\centerline{Figure 2:~{\small Continued.}}
 \label{fD}
\end{figure*}

\begin{figure*}
\hskip -4cm
    \includegraphics[width=1.2\textwidth,angle=-90]{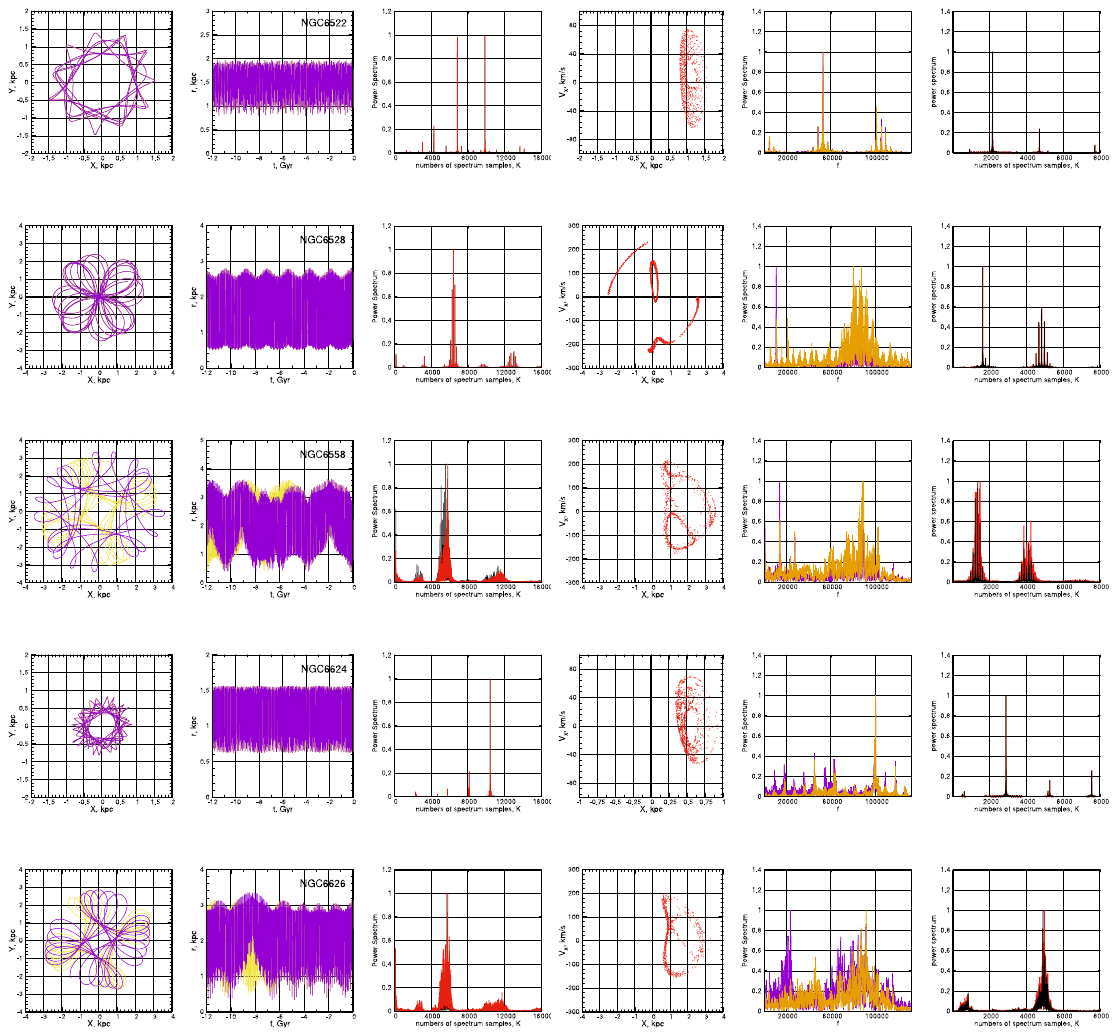}
\centerline{Figure 2:~{\small Continued.}}
 \label{fD}
\end{figure*}

\begin{figure*}
\hskip -4cm
    \includegraphics[width=1.2\textwidth,angle=-90]{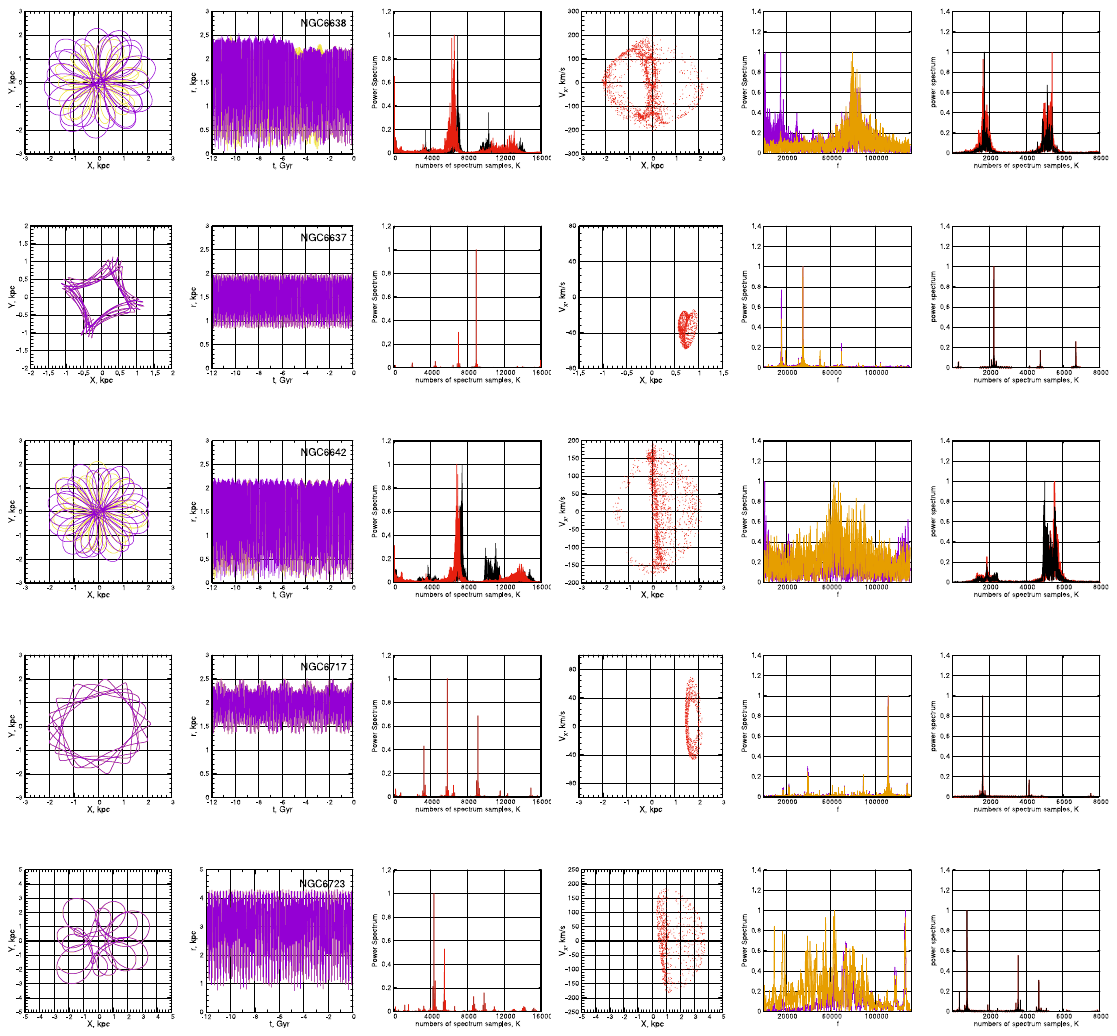}
\centerline{Figure 2:~{\small Continued.}}
 \label{fD}
\end{figure*}

\begin{figure*}
\hskip -4cm
    \includegraphics[width=1.2\textwidth,angle=-90]{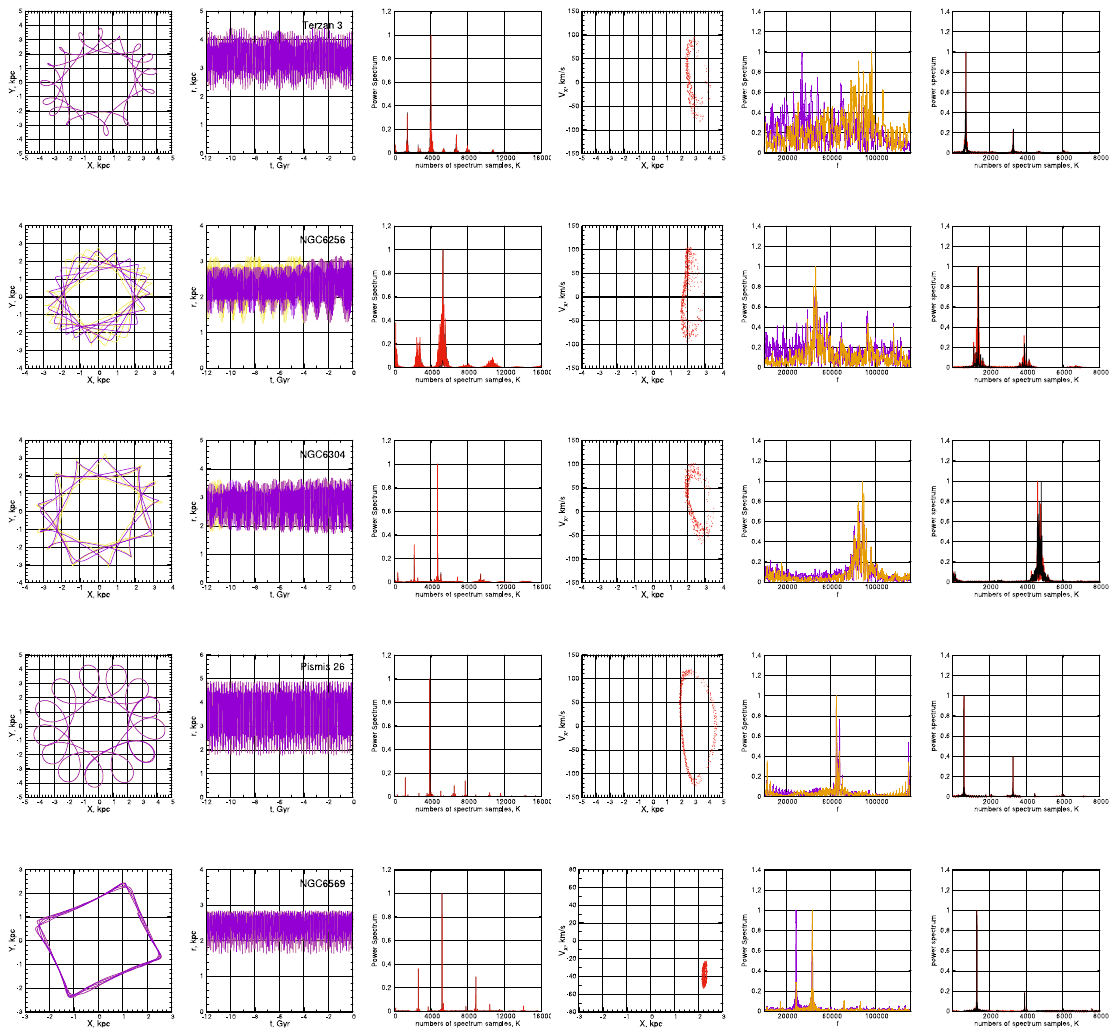}
\centerline{Figure 2:~{\small Continued.}}
 \label{fD}
\end{figure*}

\begin{figure*}
\hskip -4cm
    \includegraphics[width=1.2\textwidth,angle=-90]{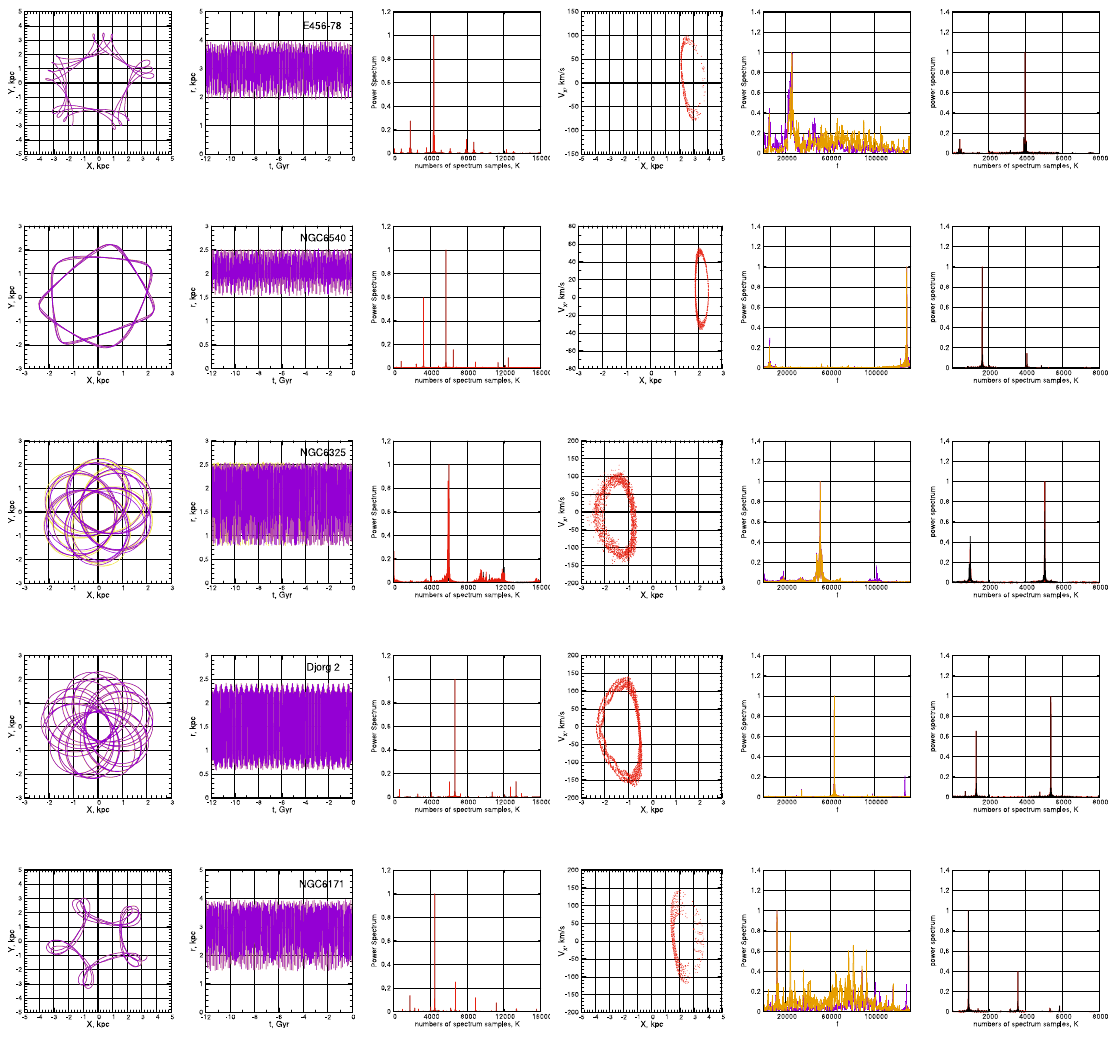}
\centerline{Figure 2:~{\small Continued.}}
 \label{fD}
\end{figure*}

\begin{figure*}
\hskip -4cm
    \includegraphics[width=1.2\textwidth,angle=-90]{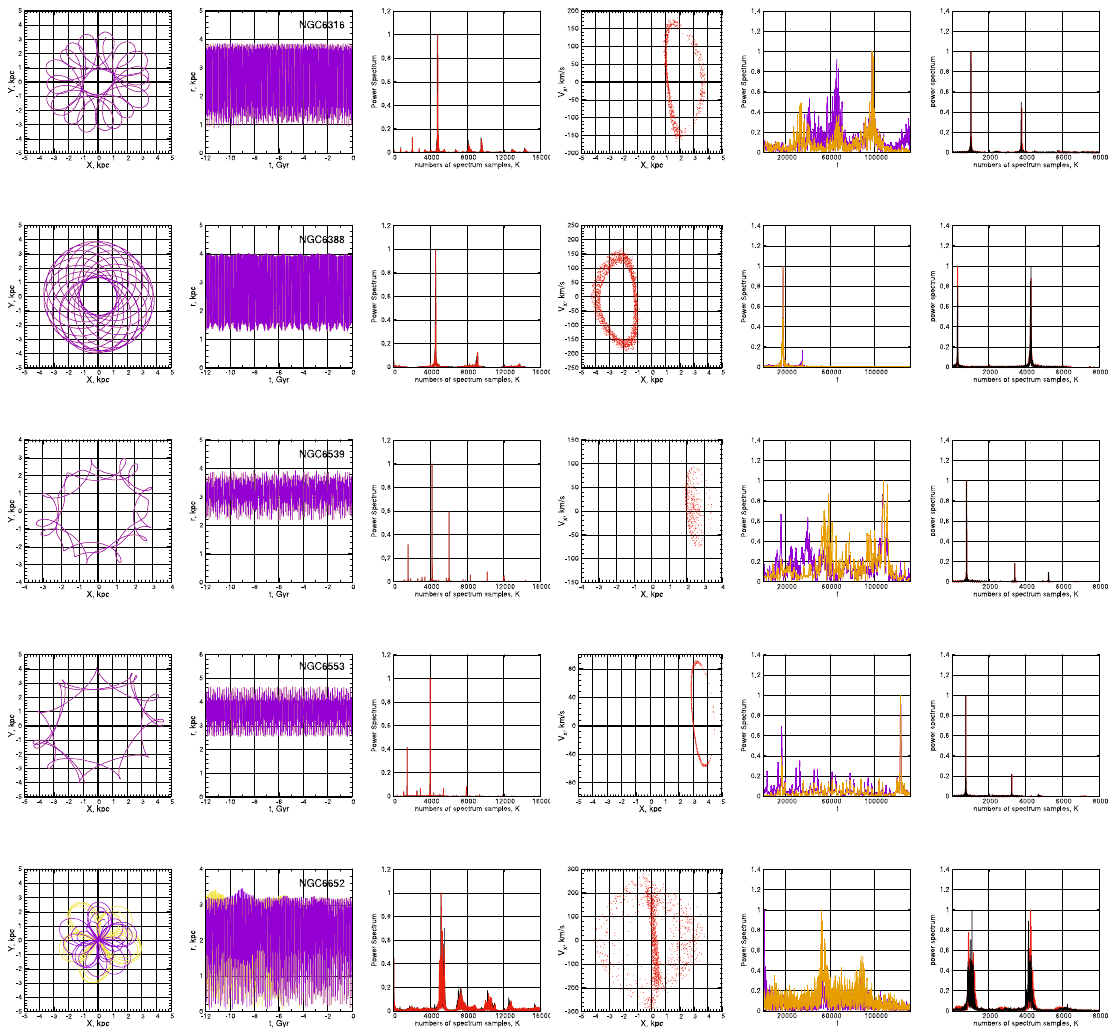}
\centerline{Figure 2:~{\small Continued.}}
 \label{fD}
\end{figure*}

\begin{figure*}
\hskip -4cm
    \includegraphics[width=1.2\textwidth,angle=-90]{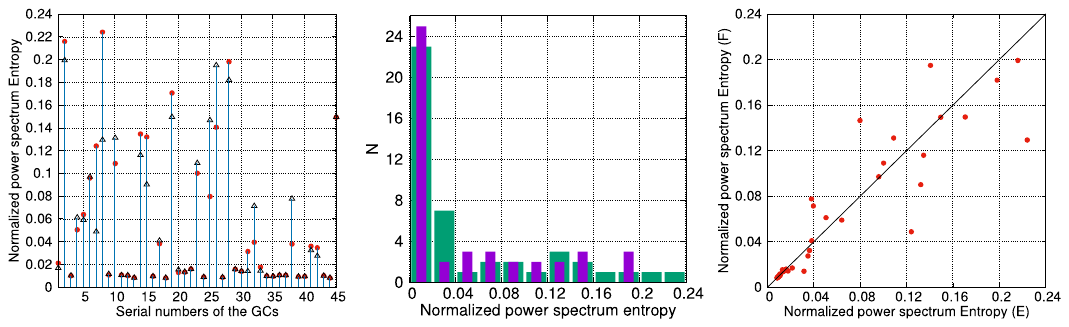}\
\vskip -6cm
\hskip 2.0cm    ({\bf a})  \hskip 4.0cm   ({\bf b}) \hskip 4cm   ({\bf c})

\caption{\small ({\bf a}) entropy values of the normalized power spectra for the 45 GCs orbits (serial numbers of the GCs are indicated on the abscissa axis) for reference (red dots) and shadow (black triangles); ({\bf b}) histograms for the entropy of normalized power spectra of reference (green) and shadow (purple) orbits of the 45 GCs; ({\bf c}) diagram comparing the entropy values of the normalized power spectra of reference (E) (abscissa axis) and shadow (F) (ordinate axis) orbits.}
 \label{fD}
\end{figure*}

 {\begin{table*}[t]                                    
 \caption[]
 {\baselineskip=1.0ex
Signs of regularity (R) or chaos (C) of the 45 GCs orbits
in the central region of the Galaxy, obtained by various methods.
  }
 \label{t:f}
 {\tiny\begin{center}\begin{tabular}{|r|l||c|c|c|c|c|c|c|}\hline
    &Name   & Entropy     &Entropy   & Entropy   &Decision &Frequency  &visual  &Estimation    \\
 N  & of the GC & for reference&for shadow & difference&by new   &drift esti-&estima- &by Poincare\\
    &       & orbits (E)  &orbits (F)& (E) and (F)& method(1)&mation (2)&tion (3)&Sections (4)  \\\hline
 1  &NGC6144 & 0.02105  & 0.01669 & 0.0044   & (C)     & (C)&  (C)    &  (C)\\\hline
 2  &E452-11 & 0.21589  & 0.19913 & 0.0168   & (C)     & (C)&  (C)    &  (C) \\\hline
 3  &NGC6266 & 0.00999  & 0.01010 & 0.0001   & (R)     & (R)&  (R)    &  (R) \\\hline
 4  &NGC6273 & 0.05026  & 0.06099 & 0.0098   & (C)     & (C)&  (C)    &  (C) \\\hline
 5  &NGC6293 & 0.06379  & 0.05878 & 0.0050   & (C)     & (C)&  (C)    &  (C) \\\hline
 6  &NGC6342 & 0.09576  & 0.09686 & 0.0011   & (C)     & (C)&  (C)    &  (C) \\\hline
 7  &NGC6355 & 0.12393  & 0.04857 & 0.0753   & (C)     & (C)&  (C)    &  (C) \\\hline
 8  &Terzan2 & 0.22402  & 0.12929 & 0.0947   & (C)     & (C)&  (C)    &  (C) \\\hline
 9  &Terzan4 & 0.01055  & 0.01143 & 0.0008   & (R)     & (R)&  (R)    &  (R) \\\hline
10  &BH229   & 0.10859  & 0.13097 & 0.0223   & (C)     & (C)&  (C)    &  (C) \\\hline
11  &Liller1 & 0.01067  & 0.01040 & 0.0002   & (R)     & (R)&  (R)    &  (R) \\\hline
12  &NGC6380 & 0.01000  & 0.01030 & 0.0003   & (R)     & (R)&  (R)    &  (R)\\\hline
13  &Terzan1 & 0.00825  & 0.00817 & 0.0001   & (R)     & (R)&  (R)    &  (R) \\\hline
14  &NGC6401 & 0.13450  & 0.11585 & 0.0186   & (C)     & (C)&  (C)    &  (C) \\\hline
15  &Pal6    & 0.13197  & 0.09002 & 0.0419   & (C)     & (C)&  (C)    &  (C) \\\hline
16  &Terzan5 & 0.00941  & 0.00944 & 0.0000   & (R)     & (R)&  (R)    &  (R)\\\hline
17  &NGC6440 & 0.03802  & 0.04076 & 0.0027   & (C)     & (C)&  (C)    &  (R) \\\hline
18  &Terzan6 & 0.00790  & 0.00812 & 0.0002   & (R)     & (R)&  (R)    &  (R) \\\hline
19  &NGC6453 & 0.17049  & 0.14939 & 0.0211   & (C)     & (C)&  (C)    &  (C) \\\hline
20  &Terzan9 & 0.01273  & 0.01526 & 0.0025   & (R)     & (R)&  (R)    &  (R) \\\hline
21  &NGC6522 & 0.01307  & 0.01339 & 0.0003   & (R)     & (R)&  (R)    &  (R) \\\hline
22  &NGC6528 & 0.01594  & 0.01571 & 0.0002   & (R)     & (R)&  (R)    &  (R) \\\hline
23  &NGC6558 & 0.09995  & 0.10890 & 0.0089   & (C)     & (C)&  (C)    &  (C) \\\hline
24  &NGC6624 & 0.00880  & 0.00883 & 0.0000   & (R)     & (R)&  (R)    &  (R) \\\hline
25  &NGC6626 & 0.07958  & 0.14641 & 0.0668   & (C)     & (C)&  (C)    &  (C) \\\hline
26  &NGC6638 & 0.14043  & 0.19477 & 0.0543   & (C)     & (C)&  (C)    &  (C) \\\hline
27  &NGC6637 & 0.00847  & 0.00866 & 0.0002   & (R)     & (R)&  (R)    &  (R) \\\hline
28  &NGC6642 & 0.19795  & 0.18179 & 0.0161   & (C)     & (C)&  (C)    &  (C) \\\hline
29  &NGC6717 & 0.01561  & 0.01555 & 0.0001   & (R)     & (R)&  (R)    &  (R) \\\hline
30  &NGC6723 & 0.01373  & 0.01397 & 0.0002   & (R)     & (R)&  (R)    &  (R)\\\hline
31  &Terzan3 & 0.03129  & 0.01390 & 0.0173   & (C)     & (R)&  (R)    &  (R) \\\hline
32  &NGC6256 & 0.03929  & 0.07113 & 0.0318   & (C)     & (C)&  (C)    &  (C)\\\hline
33  &NGC6304 & 0.01749  & 0.01415 & 0.0033   & (R)     & (C)&  (C)    &  (C)\\\hline
34  &Pismi26 & 0.00966  & 0.00963 & 0.0000   & (R)     & (R)&  (R)    &  (R) \\\hline
35  &NGC6569 & 0.00936  & 0.00931 & 0.0000   & (R)     & (R)&  (R)    &  (R) \\\hline
36  &E456-78 & 0.01041  & 0.01038 & 0.0000   & (R)     & (R)&  (R)    &  (R) \\\hline
37  &NGC6540 & 0.01034  & 0.01028 & 0.0000   & (R)     & (R)&  (R)    &  (R) \\\hline
38  &NGC6325 & 0.03775  & 0.07746 & 0.0397   & (C)     & (R)&  (C)    &  (C)\\\hline
39  &Djorg2  & 0.00913  & 0.00914 & 0,0000   & (R)     & (R)&  (R)    &  (R) \\\hline
40  &NGC6171 & 0.00922  & 0.00928 & 0.0000   & (R)     & (R)&  (R)    &  (R) \\\hline
41  &NGC6316 & 0.03595  & 0.03216 & 0.0037   & (C)     & (R)&  (R)    &  (R)\\\hline
42  &NGC6388 & 0.03460  & 0.02729 & 0.0073   & (C)     & (C)&  (R)    &  (C)\\\hline
43  &NGC6539 & 0.00991  & 0.00984 & 0.0001   & (R)     & (R)&  (R)    &  (R) \\\hline
44  &NGC6553 & 0.00805  & 0.00808 & 0.0000   & (R)     & (R)&  (R)    &  (R) \\\hline
45  &NGC6652 & 0.14926  & 0.14930 & 0.0000   & (C)     & (C)&  (C)    &  (C) \\\hline
 \end{tabular}\end{center}}\end{table*}}

\section{Conclusion}

The main result of this work is that a new simple method is proposed for determining the nature of the orbital motion (chaotic or regular) of globular clusters in the central region of the Galaxy with a radius of 3.5 kpc, which are subject to the greatest influence of the bar.

The method is based on calculating the orbital power spectrum as a function of time and calculating the entropy of the power spectrum as a measure of orbital chaos. The sample includes 45 globular clusters. To form the 6D phase space required for orbit integration, the most accurate astrometric data to date from the Gaia satellite (Vasiliev \& Baumgardt, 2021) were used, as well as new refined average distances (Baumgardt \& Vasiliev, 2021).

Orbits of globular clusters are obtained in a non-axisymmetric potential with a bar in the form of a triaxial ellipsoid embedded in an axisymmetric potential, traditionally used by us to construct orbits of globular clusters. The article provides a brief description of the potential, parameters of the axisymmetric part and bar parameters. The following, most realistic, bar parameters are adopted: mass $10^{10} M_\odot$, length of the major semi-axis 5 kpc, angle of rotation of the bar axis 25$^o$, angular velocity of rotation 40 km s$^{-1}$ kpc$^{-1}$.

A list of 23 globular clusters with regular dynamics (NGC6266, Terzan4, Liller1, NGC6380, Terzan1, Terzan5, Terzan6, Terzan9, NGC6522, NGC6528, NGC6624, NGC6637, NGC6717, NGC6723, NGC6304, Pismi26, NGC6569, E456-78, NGC6540, Djorg2, NGC6171, NGC6539, NGC6553) and a list of 22 globular clusters with chaotic dynamics (NGC6144, E452-11, NGC6273, NGC6293, NGC6342, NGC6355, Terzan2, BH229, NGC6401, Pal6, NGC6440, NGC6453, NGC6558, NGC6626, NGC6638, NGC6642, Terzan3, NGC6256, NGC6325, NGC6316, NGC6388, NGC6652).

A comparison of the classification results of the GCs obtained by the proposed method with the results obtained earlier in Bajkova et al. (2024a) by the frequency method, the Poincare section method, and also with the method of visual analysis of orbits in projection onto the galactic plane is given. The correlation of the results of the new method with those listed previously used is 82.5\%.

\subsection*{\rm \bf \normalsize References}

\setlength\parindent{-24pt}

\par

Bajkova A. T. and Bobylev V. V. (2016) Astron. Lett. 42, 567

Bajkova A. T. and Bobylev V. V. (2017) OAst 26, 72

Bajkova A. T., Carraro G., Korchagin V. I., et al. (2020) ApJ 895, 69

Bajkova A. T. and Bobylev V. V. (2022) PPulO 227, 1

Bajkova A. T., Smirnov A. A. and Bobylev V.V. (2023a) PPulO 228, 1

Bajkova A. T., Smirnov A. A. and Bobylev V.V. (2023b) PPulO 229, 1

Bajkova A. T., Smirnov A. A. and Bobylev V. V. (2023c) Ast. Bull. 78, 525

Bajkova A. T., Smirnov A. A. and Bobylev V. V. (2024a) PPulO 233, 1

Bajkova A. T., Smirnov A. A. and Bobylev V.V. (2024b) PPulO 235, 1

Baumgardt H. and Vasiliev E. (2021) MNRAS 505, 5957

Bhattacharjee P., Chaudhury S. and Kundu S. (2014) ApJ 785, 63

Bobylev V. V. and Bajkova A. T. (2016) Ast. Lett. 42, 1

Chumak O. V. (2011) Entropy and fractals in data analysis. M.-Izhevsk, 162 P.

Machado R. E. G. and Manos T. (2016) MNRAS 458, 3578

Massari D., Koppelman H. H. and Helmi A. (2019) A\&A 630, L4

Miyamoto M. and Nagai R. (1975) PASJ 27, 533

Morbidelli A. (2014) Modern Celestial Mechanics. Aspects of Solar System Dynamics. M.-Izhevsk: Institute of Computer Science, 432 P.

Navarro J. F., Frenk C. S. and White S. D. M. (1997) ApJ 490, 493

Nieuwmunster N., Schultheis M., Sormani M., et al. (2024) arXiv2403.00761

Palous J.,  Jungwiert B. and  Kopecky J. (1993) A\&A 274, 189

Sanders J. L., Smith L., Evans N. W. and Lucas P. (2019) MNRAS 487, 5188

Sch\"onrich R., Binney J. and Dehnen W. (2010) MNRAS, 403, 1829

Smirnov A. A., Bajkova A. T. and Bobylev V. V. (2023) PPulO 228, 157

Smirnov A. A., Bajkova A. T. and Bobylev V. V. (2024) MNRAS 528, 1422

Valluri M., Debattista V. P., Quinn T. and Moore B. (2010) MNRAS 403, 525

Vasiliev E. (2019) MNRAS 484, 2832

Vasiliev E. and Baumgardt H. (2021) MNRAS 505, 5978

\end{document}